\documentclass[reprint, showpacs, superscriptaddress]{revtex4-1}
\usepackage[utf8]{inputenc}
\usepackage[english]{babel}
\usepackage{amsmath}
\usepackage{hyperref}
\usepackage{amsfonts}
\usepackage{amssymb}
\usepackage{graphicx}
\usepackage{graphics}
\usepackage{multirow}
\usepackage{natbib}
\usepackage{setspace}
\usepackage{hhline}
\usepackage{pbox}
\usepackage[left=2cm,right=2cm,top=2cm,bottom=2cm]{geometry}

\begin{document}
\title{Time reversal violating Magnetic Quadrupole Moment in heavy deformed nuclei}
\author{B.G.C. Lackenby}
\affiliation{School of Physics, University of New South Wales,  Sydney 2052,  Australia}
\author{V.V. Flambaum}
\affiliation{School of Physics, University of New South Wales,  Sydney 2052,  Australia}
\affiliation{Johannes Gutenberg-Universit\"at Mainz, 55099 Mainz, Germany}
\begin{abstract}
The existence of permanent electric dipole moments (EDMs) and  magnetic quadrupole moments (MQMs) violate both time reversal invariance ($T$) and parity ($P$). Following the $CPT$ theorem they also violate combined $CP$ symmetry. Nuclear EDMs are completely screened in atoms and molecules while interaction between electrons and MQMs creates atomic and molecular EDMs which can be  measured and used to test CP-violation theories.  Nuclear MQMs are produced by the nucleon-nucleon $T,P$-odd interaction and by nucleon EDMs.   In this work we study the effect of enhancement of the nuclear MQMs due to the nuclear quadrupole deformation. Using the Nilsson model we calculate the nuclear MQMs for deformed nuclei of experimental interest and the resultant MQM energy shift in diatomic  molecules of experimental interest  $^{173}$YbF , $^{177,179}$HfF$^+$, $^{181}$TaN, $^{181}$TaO$^+$, $^{229}$ThO and $^{229}$ThF$^+$. 
\end{abstract}

\maketitle
The observed matter-antimatter asymmetry in the universe is an important open question in modern physics. Three necessary conditions were postulated by Sakarhov\cite{Sakharov1967} including the requirement that combined charge and parity ($CP$) symmetry  is violated. While the current standard model (SM)  includes a $CP$- violating mechanism through a $CP$- violating phase in the CKM matrix \cite{KM1973} this alone is insufficient to account for the observed matter anti-matter asymmetry by several orders of magnitude (see e.g. Refs.~\cite{Sakharov1967,Farrar1993, Huet1994, Pospelov2005, Canetti2012, FS2010}). Therefore, other sources and mechanisms of $CP$- violation beyond the current SM must exist and investigating these will give insight into new physics. \\ 
\linebreak
The violation of $CP$ symmetry was first detected in the decay modes of the kaon system \cite{Christenson1964} and more recently in the $B$ meson sector \cite{Belle2001, Aaij2013} however detection of $CP$- violation in other systems has not been confirmed. By the CPT theorem a mechanism which violates combined $CP$ symmetry must also violate time-reversal ($T$) symmetry. Therefore, the existence of  permanent electromagnetic moments which violate $T$ symmetry is a promising avenue for constraining theories which  incorporate a  higher degree of $CP$- violation than the SM such as supersymmetric theories which has already been tightly constrained by current experimental limits for electric dipole moments (EDMs)\cite{Pospelov2005, Safronova2017, Chupp2018}. \\
\linebreak
$CP$- violating permanent electrodynamic moments are expected to be observed in composite particles and systems  such as atoms, nuclei and baryons and interpreted as parameters of $CP$- violating interactions in the lepton and quark-gluon sectors. In this paper we focus on the magnetic quadrupole moment (MQM) of the nucleus in particular, which is the lowest order magnetic moment that is forbidden in quantum systems by the time reversal invariance ($T$) and  parity ($P$).  For an in-depth review on symmetry violating electromagnetic moments including the MQM see Ref. \cite{GF2004, KhriplovichPNC, SFK1984, Roberts2015, KhriplovichCP, Pospelov2005}. The MQM of composite systems such as the deuteron \cite{Liu2012} have previously been investigated. The search for MQM in comparison with the electrostatic $T,P$-violating moments (EDM, Schiff and octupole moments) may have the following advantages:

\begin{itemize}
\item The nuclear EDM  in neutral atoms and molecules are completely screened \cite{Schiff1963}. The Schiff and octupole moments have an additional second power of a very small nuclear radius. The magnetic interaction is not screened. The MQM contribution to atomic EDM typically is an order of magnitude larger than the contribution of the Schiff moment and several orders of magnitude larger than the octupole contribution \cite{SFK1984, Flambaum1997}.  

\item In quadrupole deformed nuclei MQM is enhanced by an order of magnitude \cite{Flambaum1994}, therefore, the MQM contribution to atomic EDM may be two orders of magnitude larger than the Schiff moment contribution.

\item In the expression for the Schiff moment there is a partial cancellation between the first term and the second (screening) term. There is also a screening correction to the octupole moment \cite{ Flambaum1986, SFK1984, Flambaum2012}.

\item In the Hg and Xe atoms  where the most accurate measurements of atomic EDM have been performed, the valence nucleon is a neutron. Therefore, the electrostatic moments (EDM, Schiff and octupole) moments do not appear directly, they exist due to the nuclear polarization effects \cite{Flambaum1986}. Due to the screening effect and the indirect polarization origin of the Schiff moment the nuclear calculations are rather unstable.
In the case of the MQM moment both valence protons and neutrons  contribute directly, and the result is expected to be more accurate\cite{Flambaum2014}.  

\end{itemize}
 A promising method of measuring $CP$- violating moments is in diatomic molecular experiments where the effective electric field is significantly larger than those directly accessible in laboratory experiments.  There is a considerable body of work for calculating the effective electric field in diatomic molecular systems which may be experimentally viable. Both theoretical and experimental progress has been made in measuring the $T,P-$ odd effects in  YbF\cite{Hudson2011, Mosyagin1998, Quiney1998, Parpia1998, Kozlov1994, Nayak2009, Steimle2007, Abe2014}, HfF$^+$ \cite{Cossel2012, Loh2013, Petrov2007, Fleig2013, Meyer2006, Skripnikov2008Hf, Le2013, Skripnikov2017Hf, Cairncross2017}, ThO \cite{Petrov2014, Meyer2008, Skripnikov2013ThO, Skripnikov2014ThO, Titov2015ThO, Fleig2014, Denis2016, Baron2017}, ThF$^+$\cite{Loh2013, Skripnikov2015Th, Denis2015}, TaN \cite{Skripnikov2015Ta, Fleig2016TaN} and TaO$^+$ \cite{Fleig2018} particularly in relation to the nuclear Schiff moment and electron EDM.   In section \ref{sec:MQMmolecule} we present the molecular energy shift due to the nuclear MQM for these molecules.\\
 \linebreak
The collective enhancement of MQM for some  heavy deformed nuclei were estimated in \cite{Flambaum1994, Flambaum2014} where they considered the contribution using a spherical wave function basis. In this work we will use the Nilsson model of the the nucleus  which is an empirically successful single particle model which accounts for the quadrupole deformation of a nucleus by using an anisotropic oscillator potential \cite{Nilsson1955, Mottelson1955, BohrMottVol2}. In the Nilsson model the deformation breaks the degeneracy of the isotropic shell model which results in several overlapping partially filled nuclear shells containing a large number of nucleons.  Each nucleon in the Nilsson model is defined in the Nilsson basis $\left[Nn_z\Lambda\Omega\right]$ where $N$ is the principle shell number ($N = n_x + n_y + n_z$), $\Lambda$ is the projection of the orbital angular momentum on the deformation axis (chosen to be the $z$-axis) and $\Omega = \Lambda + \Sigma$ is the projection of the total angular momentum of the nucleon on the deformation axis.

To illustrate why the MQM tensor should be enhanced in quadrupole deformed nuclei let us compare it with  the EDM vector property of nuclei. The direction of the EDM of a nucleon is characterised by its angular momentum projection on the deformed nucleus axis $\Omega$. In the case of the vector properties such as EDM and magnetic moment the contributions of $\Omega$ and $-\Omega$ cancel each other and  there is no enhancement in the quadrupole deformed nuclei. For the second rank tensors such as MQM and nuclear electric quadrupole moment the contributions of $\Omega$ and $-\Omega$ double the effect. There are many nucleons in the open shells of deformed nuclei and this leads to a collective enhancement of second rank tensor properties.

In the Nilsson model we consider the nucleus in the intrinsic frame which rotates with the nucleus. However the nucleus itself rotates with respect to the fixed laboratory frame \cite{BohrMottVol2}. Due to this rotation the tensor properties transform between the intrinsic  and laboratory frame. The relationship between these two frames is \cite{BohrMottVol2}
\begin{align} \label{eq:RotationalFactor}
A^{Lab} = \dfrac{I\left(2I - 1\right)}{\left(I + 1 \right)\left(2I + 3\right)}A^{Intrinsic},
\end{align}
where $I=I_z= \left|\Omega\right|$ is the projection of total nuclear angular momentum (nuclear spin) on the symmetry axis. This expression shows that only in nuclei with spin $I > 1/2$ can we detect these second order tensor properties.

\section{MQM  Calculation}
The magnetic quadrupole moment of a nucleus due to the electromagnetic current of a single nucleon with mass $m$ is defined by the second order tensor operator \cite{SFK1984},
\begin{align} \label{eq:MQMTensor}
\begin{split}
\hat{M}_{kn}^{\nu} = \dfrac{e}{2m}\left[3\mu_{\nu}\left(r_k\sigma_n + \sigma_kr_n - \dfrac{2}{3}\delta_{kn}\hat{\boldsymbol{\sigma}}\textbf{r}\right) \right. \\
\left. + 2q_{\nu}\left(r_kl_n + l_kr_n\right)\right]
\end{split}
\end{align}
where $\nu = p,n$ for protons and neutrons respectively and, $\mu_{\nu}$ and $q_{\nu}$ are the magnetic moment and charge of the nucleon respectively. The nuclear MQM is $T$-,$P$- odd and therefore it is forbidden in the absence of nucleon EDMs and $T$-, $P$- odd nuclear forces. It is understood the shell nucleons interact with the core of the nucleus through a $P-$ and $T-$ odd potential \cite{Flambaum1994, SFK1984, KhriplovichPNC}. This results in a perturbed ``spin hedgehog'' wavefunction of a nucleon given by \cite{SFK1984, Flambaum1994},
\begin{align} \label{eq:SpinHedgehog}
\left|\psi'\right> &= \left(1 + \dfrac{\xi_{\nu}}{e}\hat{\boldsymbol{\sigma}}\hat{\boldsymbol{\nabla}}\right)\left|\psi_0\right> \\
\xi_{\nu} &\approx -2\times 10^{-21}\eta_{\nu} \ e\cdot\text{cm} \nonumber
\end{align}
where $\nu = p,n$ for protons and nucleons respectively. Here $\eta_{\nu}$ represent $T$-,$P$- odd nuclear strength constants from the $T$-,$P$- violating nuclear potential $H_{T,P} = \eta_{\nu}G_{F}/(2^{3/2}m_{\nu})(\boldsymbol{\sigma}\cdot \nabla\rho)$ and $\left|\psi_0\right>$ is the unperturbed nucleon wavefunction. Here $\rho$ is the total nuclear number density and $G_{F}$ is the Fermi weak constant.  It should be noted that we used   $T$-,$P$- odd interaction in the contact limit while the actual interaction has a finite range due to the pion exchange contribution. Another approximation used in the derivation of the Eq. (\ref{eq:SpinHedgehog}) is that the strong potential and nuclear density have similar profiles (not necessarily the  spherical one). These approximations introduce a sizeable theoretical uncertainty.  Using (\ref{eq:MQMTensor}) and (\ref{eq:SpinHedgehog}) the MQM for a single nucleon due to the $P$-, $T$- odd valence-core interaction is given by,
\begin{align}
M^{TP}  = M_{zz}^{TP} = \xi\dfrac{2}{m}\left(\mu\left<\sigma\cdot l \right> - q\left<\sigma_z l_z\right>\right) .
\end{align}
In the Nilsson basis \cite{Nilsson1955} the nucleon's total angular momentum projection onto the symmetry axis is given by $\Omega_N = \Lambda_N + \Sigma_N$, where $\Sigma_N = \pm 1/2$ is the spin projection and $\Lambda$ is the orbital angular momentum projection of the nucleon. In this basis the MQM  generated by the spin-hedgehog Eq. (\ref{eq:SpinHedgehog}) is given by,
\begin{align}
M^{TP}_{\nu} = 4\Sigma_N\Lambda_N\xi\left(\mu_{\nu} - q_{\nu}\right)\dfrac{\hbar}{m_p c}.
\end{align} 
The orbit of a permanent electric dipole moment (EDM) also  generates a contribution to the nuclear MQM, $M_{\nu}^{EDM} \propto d_{\nu}$ \cite{Khriplovich1976}. As both the proton and neutron are expected to have an EDM both will contribute to the MQM. From \cite{Flambaum2014} using a valence nucleon approach the ratio of the two contributions $M^{TP}_{\nu}/M_{\nu}^{EDM}$ is independent of the total angular momentum, $I$, of the nucleon. Therefore up to non diagonal elements of definite $I$ the ratio is the same in the Nilsson model. That is,
\begin{align}
M_{\nu}^{EDM} \approx  4\Sigma_N\Lambda_N d_{\nu}\dfrac{\hbar}{m_p c}.
\end{align}
Therefore, the MQM generated by a single nucleon is given by,
\begin{align}\label{eq:NuclearMQM} 
\begin{split}
M_{\nu} &= 4\Sigma_N\Lambda_N M_{\nu}^0 \\ 
M_{\nu}^0 &= \left[\xi\left(\mu_{\nu} - q_{\nu}\right) + d_{\nu}\right]\dfrac{\hbar}{m_p c}. 
\end{split}
\end{align}
Using the Nilsson model we can find the total MQM of the nucleus by summing up every nucleon in the open and closed shells. To find the nuclear configuration of each species we have to first identify the quadrupole deformation of the nucleus. In  odd-A nuclei there is  one unpaired nucleon which defines the nucleus' spin  and parity ($I_t^{\pi}$). Therefore we find the correct deformation factor $\delta$ of the nucleus by filling up each energy level in the Nilsson energy diagrams \cite{BohrMottVol2} such that the final configuration results in the correct nuclear spin and parity (see Ref. \cite{BF2018}).   For any odd-$A$ isotope the nuclear MQM in laboratory frame can be found  using (\ref{eq:RotationalFactor})  and (\ref{eq:NuclearMQM}) if the condition $I_t \geq 3/2$ is satisfied. The nuclear MQM  for nuclei of experimental interest are presented in  Table \ref{table:NuclearMQM}. We do not consider configuration mixing in our MQM calculations. Configuration mixing has been shown to suppress the nuclear EDM and spin matrix elements with partially filled nuclear shells \cite{Yoshinaga2010, Yoshinaga2014, Yamanaka2017}. A similar effect may appear for MQM. \\

\begin{table}
\label{table:NuclearMQM}
\begin{tabular}{ccc|ccc}
\hhline{======}
Nuclei     & $I_t^{\pi}$ &  $M$ & Nuclei & $I_t^{\pi}$ & $M$ \\
\hline
$^9$Be     & $\tfrac{3}{2}^-$ & $0M_{0}^{p} + 0.4M_{0}^{n}$ &  $^{167}$Er & $\tfrac{7}{2}^+$ & $21M_{0}^{p} +36M_{0}^{n}$\\[5pt]
$^{21}$Ne  & $\tfrac{3}{2}^+$ & $0M_{0}^{p} + 0.4M_{0}^{n}$ &  $^{173}$Yb & $\tfrac{5}{2}^-$ & $14M_{0}^{p} +26M_{0}^{n}$\\[5pt]
$^{27}$Al  & $\tfrac{5}{2}^+$ & $3M_{0}^{p} + 4.5M_{0}^{n}$ &  $^{177}$Hf & $\tfrac{7}{2}^-$ & $17M_{0}^{p} +42M_{0}^{n}$\\[5pt]
$^{151}$Eu & $\tfrac{5}{2}^+$ & $12M_{0}^{p} + 23M_{0}^{n}$   &  $^{179}$Hf & $\tfrac{9}{2}^+$ & $20M_{0}^{p} + 50M_{0}^{n}$\\[5pt]
$^{153}$Eu & $\tfrac{5}{2}^+$ & $12M_{0}^{p} + 20M_{0}^{n}$   &  $^{181}$Ta & $\tfrac{7}{2}^+$ & $19M_{0}^{p} + 45M_{0}^{n}$\\[5pt]
$^{163}$Dy & $\tfrac{5}{2}^-$ & $11M_{0}^{p} + 21M_{0}^{n}$  &  $^{229}$Th & $\tfrac{5}{2}^+$ & $13M_{0}^{p} + 27M_{0}^{n}$\\[5pt]
\hhline{======}
\end{tabular}
\caption{Total nuclear MQM for each quadrupole deformed nucleus calculated  using the Nilsson model. This table presents both the proton and neutron contributions to the total nuclear MQM in the laboratory frame. }
\end{table}

Comparing these nuclear MQMs to those presented in \cite{Flambaum2014} we see that the use of the deformed Nilsson orbitals instead of the spherical orbitals leads to a  significant increase of the results. For example, in the Hafnium isotopes $^{177}$Hf and  $^{179}$Hf the neutron contribution is enhanced by a factor of 3.  Similarly, for $^{179}$Yb the neutron contribution has doubled.  Note also that  MQMs in these heavy quadrupole deformed nuclei are an order of magnitude larger than MQM due to a valence proton   ($ \sim M_{0}^{p}$)  or neutron  ($\sim M_{0}^{n}$) in spherical nuclei.

The $T$-,$P$- odd nuclear potential which generated the MQM is dominated primarily by the neutral $\pi_0$ exchange. We can express the strength constants $\eta_{\nu}$ in the  $T$-,$P$- violating nuclear potential $H_{T,P}$  in terms of the strong $\pi NN$ coupling constant $g$ and three $T$-,$P$-odd coupling constants, corresponding to the different isotopic  channels,  $g_i$ where $i=0,1,2$. For heavy nuclei the results are the following  \cite{Dmitriev1994, SFK1984}:
\begin{align}
\eta_{n} = -\eta_{p} \approx 5\times 10^{6}g\left(\bar{g}_1 + 0.4 \bar{g}_2\ - 0.2\bar{g}_0\right) .
\end{align}
We can rewrite the contribution of both the proton and nucleon MQMs in terms of these coupling constants \cite{Flambaum1994, Vorov1995},
\begin{align} 
\begin{split}
M_{p}^0(g) &= \left[ \vphantom{\dfrac{1}{2}} g\left(\bar{g}_1 + 0.4\bar{g}_2 - 0.2\bar{g}_0\right) \right. \\
&\quad \left. + \dfrac{d_{p}}{1.2 \times 10^{-14} \ e\cdot\text{cm}}\right]3.0 \times 10^{-28} \ e\cdot\text{cm}^2 
\end{split}\label{eq:MQM_pion_p} \\
\begin{split}
M_{n}^0(g) &= \left[ \vphantom{\dfrac{1}{2}} g\left(\bar{g}_1 + 0.4\bar{g}_2 - 0.2\bar{g}_0\right) \right. \\
& \ \left. + \dfrac{d_{p}}{1.3 \times 10^{-14} \ e\cdot\text{cm}}\right]3.2 \times 10^{-28} \ e\cdot\text{cm}^2. 
\end{split} \label{eq:MQM_pion_n}
\end{align}
We can write the contributions of the $T$-,$P$-odd $\pi NN$ interaction and nucleon EDMs in terms of more fundamental $T$-,$P$- violating parameters such as the  QCD $CP$- violating parameter $\bar{\theta}$ which is the heart of the strong $CP$ problem,  or in terms of the EDMs  $d$ and chromo-EDMs $\tilde{d}$ of $u$ and $d$ quarks. They are \cite{ Crewther1979,Pospelov1999, Pospelov2005, Alexandrou2017, JLQCD, PNDME2018}:
\begin{align}
g\bar{g}_0(\bar{\theta})&= -0.37 \bar{\theta} \\
\begin{split}
g\bar{g}_0(\tilde{d}_u, \tilde{d}_d)&= 0.8\times 10^{15} \left(\tilde{d}_u - \tilde{d}_{d}\right) \ \text{cm}^{-1} \\
g\bar{g}_1(\tilde{d}_u, \tilde{d}_d)&= 4\times 10^{15} \left(\tilde{d}_u + \tilde{d}_{d}\right) \ \text{cm}^{-1}
\end{split} \label{eq:EDM_Chromo_1} \\
\begin{split}
d_{p}(d_u, d_d, \tilde{d}_u, \tilde{d}_d) &= 1.1e\left(\tilde{d}_u + 0.5\tilde{d}_{d}\right) + 0.8 d_u - 0.2d_d \\
d_{n}(d_u, d_d, \tilde{d}_u, \tilde{d}_d) &= 1.1e\left(\tilde{d}_d + 0.5\tilde{d}_{u}\right) - 0.8 d_d + 0.2d_u
\end{split} \label{eq:EDM_Chromo_2}
\end{align}
where the chromo-EDM contributions in eqs. (\ref{eq:EDM_Chromo_1}) and (\ref{eq:EDM_Chromo_2}) arise from the Peccei-Quinn mechanism \cite{Peccei1977, Pospelov2005}. 
The corresponding substitutions give the following results for the dependence on $\tilde{\theta}$ of  proton and neutron MQM contributions:
\begin{align}
\begin{split}
M_{p}^0(\bar{\theta}) = 1.9 \times 10^{-29}\bar{\theta} \ e\cdot\text{cm}^2 \\
M_{n}^0(\bar{\theta}) = 2.5 \times 10^{-29}\bar{\theta} \ e\cdot\text{cm}^2.
\end{split}
\end{align}
The dependence  on the up and down quark EDMs is
\begin{align} 
\begin{split}
M_{p}^0(\tilde{d}_u - \tilde{d}_d) = 1.2 \times 10^{-12}(\tilde{d}_u - \tilde{d}_d) \ e\cdot\text{cm} \\
M_{n}^0(\tilde{d}_u - \tilde{d}_d) = 1.3 \times 10^{-12}(\tilde{d}_u - \tilde{d}_d) \ e\cdot\text{cm}. 
\end{split}
\end{align}
While there have been more sophisticated treatments of the  $\pi NN$ interaction with respect to $\bar{\theta}$\cite{Vries2015, Engel2013, Yamanaka2017, Chupp2018} and the quark chromo-EDMs \cite{Fuyuto2013, Seng2018, Engel2013, Yamanaka2017, Chupp2018} the values used above are within the accuracy of our model.  
\section{MQM energy shift in diatomic molecules} \label{sec:MQMmolecule}
Direct measurement of the nuclear MQM is unfeasible and a more indirect method is required. As mentioned above the use of neutral molecular systems is promising as the nuclear MQM will interact with the internal electromagnetic field. Molecules in particular present a lucrative option due to existence of very close paired levels of opposite parity, the  $\Omega$-doublet - see e.g.  \cite{Flambaum2014}. For highly polar molecules consisting of a heavy and light nucleus (for example, Th and O) the effect of MQM is $\sim Z^2$, therefore it is calculated for the heavier nucleus. The Hamiltonian of diatomic paramagnetic molecule including the $T, P-$ odd nuclear moment effects is given by \cite{SFK1984,Kozlov1995}:
\begin{align}
H = W_d d_e \mathbf{S}\cdot\mathbf{n} + W_{Q}\dfrac{Q}{I}\mathbf{I}\cdot\mathbf{n} - \dfrac{W_{M}M}{2I(2I -1)}\mathbf{S}\hat{\mathbf{T}}\mathbf{n},
\end{align}
where $d_e$ is the electron EDM, $Q$ is the nuclear Schiff moment, $M$ is the nuclear MQM, $\mathbf{S}$ is the electron spin, $\mathbf{n}$ is the symmetry axis of the molecule, $\hat{\mathbf{T}} $ is the second rank tensor operator characterised by the nuclear spins $T_{ij} = I_iI_j + I_jI_i - \tfrac{2}{3}\delta_{ij}I(I + 1)$  and  $W_d$, $W_Q$ and $W_M$ are fundamental parameters for each interaction which are dependent on the particular molecule. We have omitted the $P$-,$T$- odd electron-nucleon interaction terms which are presented e.g. in reviews \cite{Safronova2017,GF2004}. 
These parameters $W_d$, $W_Q$ and $W_M$
are related to the electronic molecular structure of the state. For each molecule there is an effective field for each fundamental parameter, these effective fields are calculated using many-body methods for electrons close to the heavy nucleus \cite{Flambaum2014}.  For the nuclear MQM we are interested only in $W_M$ which has been calculated for molecules YbF \cite{Kozlov1995}, HfF$^{+}$\cite{Skripnikov2017}, TaN \cite{Skripnikov2015Ta, Fleig2016TaN}, TaO$^+$\cite{Fleig2018}, ThO \cite{Skripnikov2014ThO} and ThF$^+$\cite{Skripnikov2015Th}. 
Using these vales we present the results for the energy shifts in molecules induced by MQM in terms of $CP-$ violating parameters $\bar{\theta}$, $d_p$  and $(\tilde{d}_u -\tilde{d}_d)$ in Table \ref{table:MQMMoleculeShift}. 

\begin{table}
\footnotesize
\begin{tabular}{c c c c c c c}
\hline
\hline
 & & & $|W_M|$ & \multicolumn{3}{c}{$|W_M M S|$ ($\mu$Hz)} \\[5pt]
Molecule & $I_t^{\pi}$ & State & \pbox{20cm}{$10^{39}$ $\mu$Hz/\\ 
 $e\cdot$cm$^{2}$}  & \pbox{20cm}{$10^{25}d_p$/ \\
 $e\cdot$ cm} & \pbox{20cm}{$10^{10} \bar{\theta}$} & \pbox{20cm}{$10^{27}(\tilde{d}_{u} - \tilde{d}_d)$/ \\
 cm} \\[5pt]
\hline
$^{173}$YbF & $\tfrac{5}{2}^-$ & $^{2}\Sigma_{1/2}$ & 2.1\cite{Kozlov1995} & 37 & 96 & 53 \\[5pt]
$^{177}$HfF$^+$  & $\tfrac{7}{2}^-$ & $^3\Delta_1$ & 0.494\cite{Skripnikov2017} & 21 & 68 & 37 \\[5pt]
$^{179}$HfF$^+$ & $\tfrac{9}{2}^+$ & $^3\Delta_1$ & 0.494\cite{Skripnikov2017} & 25 & 81 & 44 \\[5pt]
$^{181}$TaN & $\tfrac{7}{2}^+$ & $^3\Delta_1$ & 1.08\cite{Skripnikov2015Ta} & 51 & 159 & 87 \\[5pt]
$^{181}$TaO$^+$ & $\tfrac{7}{2}^+$ & $^3\Delta_1$  & 0.45\cite{Fleig2018} & 21 & 66 & 36 \\[5pt]
$^{229}$ThO & $\tfrac{5}{2}^+$ & $^3\Delta_1$ & 1.10\cite{Skripnikov2014ThO} & 35 & 102 & 56\\[5pt]
$^{229}$ThF$^+$ & $\tfrac{5}{2}^+$ & $^3\Delta_1$ &0.88\cite{Skripnikov2015Th} & 28 & 81 & 45\\[5pt] 
\hline 
\hline
\end{tabular}
\caption{Frequency shifts due to the MQM interaction with the electron magnetic  field of the molecules. We present the energy shifts in terms of the $CP-$ violating parameters of interest. These are the strong $CP-$ term in QCD $\bar{\theta}$, the permanent EDM of the proton $d_p$ and the difference of quark chromo-EDMs $(\tilde{d}_{u} - \tilde{d}_d)$.\label{table:MQMMoleculeShift}}
\end{table}
The MQM molecular energy shifts for HfF$^+$, TaN, TaO$^{+}$ and ThO were calculated in Refs. \cite{Skripnikov2017Hf}, \cite{Skripnikov2015Ta}, \cite{Fleig2018} and \cite{Skripnikov2014ThO} respectively. They used the MQM calculated in  the spherical basis method outlined in \cite{Flambaum2014} and represent the shifts in fundamental $T$-,$P$- odd parameters as in Table \ref{table:MQMMoleculeShift}. Using the Nilsson model, the MQM energy shifts are larger for TaN, TaO$^+$ and ThO molecules by a factor of 2 however for $^{177}$HfF$^+$ the values of the two models are similar. Using the currents limits on the CP-violating parameters \cite{Swallows2013} $|d_p| < 8.6 \times 10^{-25}$~$e\cdot$cm, $\bar{\theta} < 2.4 \times 10^{10}$ and $\tilde{d}_{u} - \tilde{d}_d < 6\times 10^{-27}$~cm the respective MQM energy shifts ($\left|W_M M S\right|$) in $^{229}$ThO are $<300$~$\mu$Hz, $<250$~$\mu$Hz and $340$~$\mu$Hz. The $^{232}$ThO molecule has recently been used to set new limits on the electron EDM with a factor of 12 improvement in accuracy of 80 $\mu$Hz\cite{ACME2014,ACME2018}. As $^{232}$Th has an even number of nucleons there is no spectroscopic nuclear MQM. Therefore in principle, if a similar experiment is possible with $^{229}$ThO future measurements should improve constraints on nuclear $CP$- violating interactions. It is interesting to find the minimal SM prediction for the energy shifts which comes solely from the CKM matrix. Using eqs. (\ref{eq:MQM_pion_p}) and (\ref{eq:MQM_pion_n}), the lower limit on the CKM nucleon EDM $d_{p}^{\text{{\tiny CKM}}} = -d_{n}^{\text{{\tiny CKM}}} \approx 1 \times 10^{-32}$~$e\cdot$cm\cite{Seng2015} and the strengths of the $CP-$odd pion  nucleon couplings in the CKM model $g\bar{g}_0 \approx -1.6 \times 10^{-16}$, $g\bar{g}_1 \approx -1.8 \times 10^{-16}$ and $g\bar{g}_2 \approx 4.7\times 10^{-20}$ \cite{Yamanaka2016} we find $|M_{p}^{0,\text{{\tiny CKM}}}|\approx |M_{n}^{0,\text{{\tiny CKM}}}| \approx 4.5 \times 10^{-44} \ e\cdot$cm$^2$. This corresponds to an energy shift of $|W_MMS| \approx 1$~nHz in $^{229}$ThO due to the MQM which is 4 orders of magnitude lower than the current accuracy. Results for other molecules in Table \ref{table:MQMMoleculeShift} are similar.

\section{Conclusion}
In this work we present a novel method for calculating the nuclear MQM for any nuclei that satisfy the angular momentum condition $I_t \geq 3/2$. In heavy nuclei with large quadrupole deformations there is an enhancement of the nuclear MQM  due to the collective effect of partially filled nucleon shells and therefore these nuclei present an opportunity for detecting and measuring $T$-,$P$- violating effects in the hadronic sector. The molecular systems which have been used to study the electron EDM with promising results are also excellent candidates for measuring the nuclear MQM  \cite{Skripnikov2017, Skripnikov2014ThO}. With increasing experimental capabilities  in paramagnetic molecular systems the possibility of measure these $T$-,$P$- violating effects is attractive. The nuclear MQM's and MQM molecular energy shifts presented in this work may allow experimentalists either detect or constrain the limits of fundamental $T$-,$P$- violating nucleon EDM ($d_p$), strong $CP$ parameter ($\bar{\theta}$) and chromo-EDMs $(\tilde{d}_{u} - \tilde{d}_d) $. \\

This work is supported in part by the Australian Research Council, the Gutenberg Fellowship and by the National Science Foundation under grant No. NSF PHY11-25915. V.F. is grateful to Kavli Institute for Theoretical Physics at Santa Barbara for  hospitality.
\bibliographystyle{apsrev4-1}
\bibliography{MQM_Nilsson}
\end{document}